\documentclass[12pt]{article}
\usepackage{fancyhdr}

\usepackage{mathrsfs}
\usepackage[T1]{fontenc}
\usepackage{mathpazo}
\usepackage{setspace}
\usepackage{amsfonts}
\usepackage{amssymb}
\usepackage{amsmath}
\usepackage{epsfig}
\usepackage{latexsym}
\usepackage{color}
\usepackage{graphicx}
\usepackage{nicefrac}
\usepackage[latin1]{inputenc}
\usepackage{slashed}
\usepackage{multirow}
\usepackage{comment}
\usepackage{soul}
\usepackage{hyperref}
\usepackage{cite}

 \usepackage{empheq}

    \newlength\fsep
    \newsavebox\widebox
\newenvironment{mathbox}
    {\par\vskip\fsep\noindent%
     \begin{lrbox}{\widebox}%
     \begin{minipage}{\textwidth-\fsep}%
    }{\vskip\fsep\end{minipage}\end{lrbox}
      \framebox{\usebox\widebox}%
     }

\usepackage[titletoc]{appendix}

\def\hybrid{\topmargin -20pt    \oddsidemargin 0pt
        \headheight 0pt \headsep 0pt
        \textwidth 6.25in       
        \textheight 9.25in       
        \marginparwidth .875in
        \parskip 5pt plus 1pt   \jot = 1.5ex}

\hybrid

\def\baselinestretch{1.2}

\catcode`\@=11

\def\marginnote#1{}
\newcount\hour
\newcount\minute
\newtoks\amorpm
\hour=\time\divide\hour by60
\minute=\time{\multiply\hour by60 \global\advance\minute by-\hour}
\edef\standardtime{{\ifnum\hour<12 \global\amorpm={am}%
        \else\global\amorpm={pm}\advance\hour by-12 \fi
        \ifnum\hour=0 \hour=12 \fi
        \number\hour:\ifnum\minute<10 0\fi\number\minute\the\amorpm}}
\edef\militarytime{\number\hour:\ifnum\minute<10 0\fi\number\minute}

\def\draftlabel#1{{\@bsphack\if@filesw {\let\thepage\relax
   \xdef\@gtempa{\write\@auxout{\string
      \newlabel{#1}{{\@currentlabel}{\thepage}}}}}\@gtempa
   \if@nobreak \ifvmode\nobreak\fi\fi\fi\@esphack}
        \gdef\@eqnlabel{#1}}
\def\@eqnlabel{}
\def\@vacuum{}
\def\draftmarginnote#1{\marginpar{\raggedright\scriptsize\tt#1}}

\def\draft{\oddsidemargin -.5truein
        \def\@oddfoot{\sl preliminary draft \hfil
        \rm\thepage\hfil\sl\today\quad\militarytime}
        \let\@evenfoot\@oddfoot \overfullrule 3pt
        \let\label=\draftlabel
        \let\marginnote=\draftmarginnote
   \def\@eqnnum{(\theequation)\rlap{\kern\marginparsep\tt\@eqnlabel}%
\global\let\@eqnlabel\@vacuum}  }

\def\preprint{\twocolumn\sloppy\flushbottom\parindent 2em
        \leftmargini 2em\leftmarginv .5em\leftmarginvi .5em
        \oddsidemargin -.5in    \evensidemargin -.5in
        \columnsep .4in \footheight 0pt
        \textwidth 10.in        \topmargin  -.4in
        \headheight 12pt \topskip .4in
        \textheight 6.9in \footskip 0pt
        \def\@oddhead{\thepage\hfil\addtocounter{page}{1}\thepage}
        \let\@evenhead\@oddhead \def\@oddfoot{} \def\@evenfoot{} }

\def\numberbysection{\@addtoreset{equation}{section}
        \def\theequation{\thesection.\arabic{equation}}}

\def\underline#1{\relax\ifmmode\@@underline#1\else
        $\@@underline{\hbox{#1}}$\relax\fi}

\def\titlepage{\@restonecolfalse\if@twocolumn\@restonecoltrue\onecolumn
     \else \newpage \fi \thispagestyle{empty}\c@page\z@
        \def\thefootnote{\fnsymbol{footnote}} }

\def\endtitlepage{\if@restonecol\twocolumn \else \newpage \fi
        \def\thefootnote{\arabic{footnote}}
        \setcounter{footnote}{0}}  

\catcode`@=12
\relax

\def\figcap{\section*{Figure Captions\markboth
        {FIGURECAPTIONS}{FIGURECAPTIONS}}\list
        {Figure \arabic{enumi}:\hfill}{\settowidth\labelwidth{Figure
999:}
        \leftmargin\labelwidth
        \advance\leftmargin\labelsep\usecounter{enumi}}}
 \relax
\def\tablecap{\section*{Table Captions\markboth
        {TABLECAPTIONS}{TABLECAPTIONS}}\list
        {Table \arabic{enumi}:\hfill}{\settowidth\labelwidth{Table
999:}
        \leftmargin\labelwidth
        \advance\leftmargin\labelsep\usecounter{enumi}}}
 \relax
\def\reflist{\section*{References\markboth
        {REFLIST}{REFLIST}}\list
        {[\arabic{enumi}]\hfill}{\settowidth\labelwidth{[999]}
        \leftmargin\labelwidth
        \advance\leftmargin\labelsep\usecounter{enumi}}}
 \relax
\makeatletter
\newcounter{pubctr}
\def\publist{\@ifnextchar[{\@publist}{\@@publist}}
\def\@publist[#1]{\list
        {[\arabic{pubctr}]\hfill}{\settowidth\labelwidth{[999]}
        \leftmargin\labelwidth
        \advance\leftmargin\labelsep
        \@nmbrlisttrue\def\@listctr{pubctr}
        \setcounter{pubctr}{#1}\addtocounter{pubctr}{-1}}}
\def\@@publist{\list
        {[\arabic{pubctr}]\hfill}{\settowidth\labelwidth{[999]}
        \leftmargin\labelwidth
        \advance\leftmargin\labelsep
        \@nmbrlisttrue\def\@listctr{pubctr}}}
 \relax
\makeatother
\newskip\humongous \humongous=0pt plus 1000pt minus 1000pt

\newif\ifdtup

\relax

\def\be{\begin{equation}}
\def\ee{\end{equation}}
\def\ba{\begin{eqnarray}}
\def\ea{\end{eqnarray}}

\def\del{\partial}

\def\r{\rho}

\def\g{\gamma}

\def\d{\delta}
\def\D{\Delta}

\def\m{\mu}

\def\l{\lambda}

\def\s{\sigma}
\def\S{\Sigma}

\def\no{\noindent}

\def\qq{\qquad}

\def\IR{\relax{\rm I\kern-.18em R}}

\def \J {{\bar J} }

\def \ov {\over}

\def\IR{\relax{\rm I\kern-.18em R}}
\def\IL{\relax{\rm I\kern-.18em L}}

\def\inv{^{\raise.15ex\hbox{${\scriptscriptstyle -}$}\kern-.05em 1}}

\begin{document}

\renewcommand{\theequation}{\thesection.\arabic{equation}}
\csname @addtoreset\endcsname{equation}{section}

\newcommand{\beq}{\begin{equation}}
\newcommand{\eeq}[1]{\label{#1}\end{equation}}
\newcommand{\ber}{\begin{equation}}
\newcommand{\eer}[1]{\label{#1}\end{equation}}
\newcommand{\eqn}[1]{(\ref{#1})}
\begin{titlepage}
\begin{center}


${}$
\vskip .2 in

{\large\bf $\lambda$-deformations of left--right asymmetric CFTs}

\vskip 0.4in

{\bf George Georgiou,$^1$\ Konstantinos Sfetsos}$^{2}$\ and\ {\bf Konstantinos Siampos}$^{3}$
\vskip 0.1in

\vskip 0.1in
{\em
${}^1$Institute of Nuclear and Particle Physics,\\ National Center for Scientific Research Demokritos,\\
Ag. Paraskevi, GR-15310 Athens, Greece
}
\vskip 0.1in

 {\em
${}^2$Department of Nuclear and Particle Physics,\\
Faculty of Physics, National and Kapodistrian University of Athens,\\
Athens 15784, Greece\\
}
\vskip 0.1in

{\em ${}^3${Albert Einstein Center for Fundamental Physics,\\
Institute for Theoretical Physics / Laboratory for High-Energy Physics,\\
University of Bern,
Sidlerstrasse 5, CH3012 Bern, Switzerland
}}

\vskip 0.1in

{\footnotesize \texttt georgiou@inp.demokritos.gr, ksfetsos@phys.uoa.gr, siampos@itp.unibe.ch}


\vskip .5in
\end{center}

\centerline{\bf Abstract}

\no
We compute the all-loop anomalous dimensions of current and primary field operators in
deformed current algebra theories based on a general semi-simple group,
but with different (large) levels for the left and right sectors.
These theories, unlike their equal level counterparts,
possess a new non-trivial fixed point in the IR. 
By computing  the exact in $\lambda$ two- and three-point functions for these operators
we deduce their OPEs and their equal-time commutators.
Using these we argue on the nature of the CFT at the IR fixed point.
The associated to the currents Poisson
brackets are a two-parameter deformation of the canonical structure of the isotropic PCM.

\vskip .4in
\noindent
\end{titlepage}
\vfill
\eject

\newpage

\tableofcontents

\noindent

\def\baselinestretch{1.2}
\baselineskip 20 pt
\noindent


\setcounter{equation}{0}
\section{Introduction}
\renewcommand{\theequation}{\thesection.\arabic{equation}}

We are interested in a two-dimensional conformal field theory (CFT) which possesses two independent
current algebras generated by $J_a(z) $ and $\bar J_a(\bar z)$ which are holomorphic and anti-holomorphic,
respectively.
The index $a$ takes values $a=1,2,\dots , \dim G$, where $G$ is the non-Abelian group of the CFT.
The singular part of the holomorphic currents OPE reads
\cite{Witten:1983ar,Knizhnik:1984nr}
\begin{equation}
\label{OPE}
J_{a}(z)J_{b}(w)=\frac{\delta_{ab}}{(z-w)^2}+\frac{f_{abc}}{\sqrt{k_L}} {J_c(w)\ov  z-w}+\cdots\ ,
\end{equation}
where the level $k_L$ is a positive integer.
A similar expression holds for the OPEs between the antiholomorphic currents $\bar J_a(\bar z)$, but with
a different level $k_R$. Of course, the OPE $J_{a}(z)\bar{J}_b(\bar w)$ is regular.

\no
We now consider this $G_{k_L}\times G_{k_R}$ current algebra
theory perturbed with a classically marginal operator bilinear in the currents.
In the Euclidean regime the action reads
\be
S= S_{{\rm CFT};k_L,k_R} -{\l\ov \pi}\int \text{d}^2z\, J_a(z) \bar J_a(\bar z) \ .
\label{thir}
\ee
An explicit example of such an action is the bosonized non-Abelian Thirring model action (for a general
discussion, see \cite{Dashen:1974gu,Karabali:1988sz}), namely the WZW model two-dimensional
CFT perturbed by the above current bilinear. In that case, however
the two levels are equal, i.e. $k_L=k_R$.
In what follows we will not need an explicit form for the action $S_{{\rm CFT};k_L,k_R}$ of the unperturbed CFT.

The unequal level case has some very interesting features, not present when $k_L=k_R$.
The basic one is that under RG flow the theory reaches a new fixed point in the IR which lies within the perturbative domain.
In contrast, in the equal level case the RG flow drives the theory to a strong coupling regime.
In recent years, a lot of progress has been made in the left--right symmetric case.
This started with the construction \cite{Sfetsos:2013wia} of  the all-loop effective action corresponding to \eqn{thir} .
The corresponding   $\s$-model was shown to be  integrable \cite{Sfetsos:2013wia}. Subsequently, the extension of this construction to cosets \cite{Sfetsos:2013wia,Hollowood:2014rla,Hollowood:2014qma} and to supergroups \cite{Hollowood:2014rla,Hollowood:2014qma} took place,
while the computation of the general RG flow equations using the effective action and gravitational methods was performed in \cite{Itsios:2014lca,Sfetsos:2014jfa}.
In parallel developments, these models, named generically as $\l$-deformed, were embedded in specific
cases to supergravity \cite{Sfetsos:2014cea,Demulder:2015lva,Borsato:2016zcf,Chervonyi:2016ajp,Chervonyi:2016bfl}.
In addition, a classical relation to $\eta$-deformations \cite{Klimcik:2002zj,Klimcik:2008eq,Klimcik:2014}
and in \cite{Delduc:2013fga,Delduc:2013qra,Arutyunov:2013ega}, via Poisson-Lie T-duality
\cite{KS95a} and appropriate analytic continuations was uncovered in \cite{Vicedo:2015pna,Hoare:2015gda,Sfetsos:2015nya,Klimcik:2015gba,Klimcik:2016rov}.
More recently the computation of the all-loop anomalous dimensions
and of current and primary field operators was performed \cite{Georgiou:2016iom, Georgiou:2015nka}.
In view of the very interesting and totally different behaviour under the RG-flow that we will shortly see,
it is natural to push similar investigations in the case of left--right asymmetry, as well.

\no
In our computation we will make use of the basic
two- and three-point functions at the CFT point given for the holomorphic currents by
\begin{equation}
\langle J_a(z_1) J_b(z_2)\rangle ={\delta_{ab}\ov z_{12}^2}\ ,\qq
\langle J_a(z_1) J_b(z_2) J_c(z_3)\rangle ={1\ov \sqrt{k_L}} {f_{abc}\ov z_{12} z_{13} z_{23}}\ ,
\label{23point}
\end{equation}
where we employ the general notation $z_{ij}= z_i-z_j$.
Similar expressions hold for the antiholomorphic currents with the replacement of $k_L$ by $k_R$.
Mixed correlators involving holomorphic and anti-holomorphic currents vanish.
In addition, in order to compute higher order correlation functions,
we will employ the Ward identity
\ba
&& \langle J_a(z) J_{a_1}(z_1) J_{a_2}(z_2)\cdots J_{a_n}(z_n)\rangle =
{1\ov \sqrt{k_L}} \sum_{i=1}^n {f_{a a_i b}\ov z-z_i}
\langle  J_{a_1}(z_1) J_{a_2}(z_2)\cdots J_b(z_i) \cdots J_{a_n}(z_n)\rangle
\nonumber\\
&&
\phantom{xxxx}
+  \sum_{i=1}^n {\d_{a a_i}\ov (z-z_i)^2}
\langle  J_{a_1}(z_1) J_{a_2}(z_2)\cdots J_{a_{i-1}}(z_{i-1})
J_{a_{i+1}}(z_{i+1})\cdots J_{a_n}(z_n)\rangle\
\label{WardIdent}
\ea
and a similar one for the anti-holomorphic sector.

\no
The CFT theory contains also affine primary fields $\Phi_{i,i'}(z,\bar z)$ transforming in irreducible representations $R$ and $R'$
of the Lie algebra for $G$, with Hermitian matrices $t_a$ and $\tilde t_a$.
Under the action of the currents they transform as \cite{Knizhnik:1984nr}
\begin{equation}
\label{jjj}
\begin{split}
& J_a(z) \Phi_{i,i'}(w,\bar w) = -{1\ov \sqrt{k_L}} {(t_a)_i{}^j \Phi_{j,i'}(w,\bar w)\ov z-w}\ ,
\\
&\bar J_a(\bar z) \Phi_{i,i'}(w,\bar w) = {1\ov \sqrt{k_R}} {(\tilde t_a)^{j'}{}_{i'} \Phi_{i,j'}(w,\bar w)\ov \bar z- \bar w}\ ,
\end{split}
\end{equation}
where $[t_a,t_b]=f_{abc}t_c$ and $[\tilde t_a,\tilde t_b]=f_{abc}\tilde t_c$, with
$i=1,2,\dots , \dim R$ and $i'=1,2,\dots , \dim R'$.
These fields are also Virasoro primaries with holomorphic and
antiholomorphic dimensions given by \cite{Knizhnik:1984nr}
\begin{equation}
\D_R = {c_R\ov 2k_L+ c_G}\ ,\qq \bar \D_{R'} = {c_{R'}\ov 2k_R+ c_G}\ ,
\label{ddcft}
\end{equation}
where $c_R$, $c_{R'}$ and $c_G$ are the quadratic Casimir operators in the representations $R$ and $R'$
and in the adjoint representation.
They are defined as
\begin{equation}
(t_a t_a)_i{}^j = c_R \delta_i{}^j \ ,\qq (\tilde t_a \tilde t_a)_{i'}{}^{j'}
= c_{R'} \delta_{i'}{}^{j'}\ ,\qq  f_{acd} f_{bcd} = - c_G \delta_{ab} \ .
\end{equation}
The Virasoro central charges are
\be
C_L = {2 k_L\dim G\ov 2 k_L + c_G}\ ,\qq C_R = {2 k_R\dim G\ov 2 k_R + c_G}\ .
\label{centraal}
\ee
Of particular importance, especially in considerations in subsection (2.4), will be the adjoint representation $\Phi_{a,b}$
for which the representations matrices are  $(t_a)_{bc}=(\tilde t_a)_{bc} =- f_{abc}$.
Then we have that
\be
\begin{split}
&
J_a(z) \Phi_{b,c}(w,\bar w)= {1\ov \sqrt{k_L}} {f_{abd} \Phi_{d,c}(w,\bar w)\ov z-w}\ ,
\label{adjjj}
\\
&
\bar J_a (\bar z) \Phi_{b,c}(w,\bar w)= {1\ov \sqrt{k_R}} {f_{acd} \Phi_{b,d}(w,\bar w)\ov \bar z-\bar w}\ .
\end{split}
\ee

\no
The two-point correlators for the affine primaries are
\begin{equation}
\langle \Phi^{(1)}_{i,i'}(z_1,\bar z_1) \Phi^{(2)}_{j,j'}(z_2,\bar z_2)\rangle
={\delta_{ij}\,\delta_{i'j'}\ov z_{12}^{2\D_R} \ \bar z_{12}^{2 \bar \D_{R'}}}\ ,
\end{equation}
where the superscripts denote the different representations the primaries belong to.
In addition, the mixed three-point functions involving one current are given by
\begin{equation}
\langle J_a(x_3) \Phi^{(1)}_{i,i'}(x_1,\bar x_1) \Phi^{(2)}_{j,j'}(x_2,\bar x_2)\rangle
= {1\ov \sqrt{k_L}} {(t_a\otimes \mathbb{I}_{R'})_{ii',jj'}\ov x_{12}^{2 \D_R}\ \bar x_{12}^{2 \bar \D_{R'}}}
\left({1\ov x_{13}}-{1\ov x_{23}}\right)
\label{JJf1}
\end{equation}
and
\begin{equation}
\langle \bar J_a(\bar x_3) \Phi^{(1)}_{i,i'}(x_1,\bar x_1) \Phi^{(2)}_{j,j'}(x_2,\bar x_2)\rangle
= - {1\ov \sqrt{k_R}} { (\mathbb{I}_{R}\otimes \tilde t_a^*)_{ii',jj'}\ov x_{12}^{2 \D_R}\ \bar x_{12}^{2 \bar \D_{R'}}}
\left({1\ov \bar x_{13}}-{1\ov \bar x_{23}}\right)\ .
\label{JJf2}
\end{equation}
These correlators are non-vanishing as long as the representations $R$
and $R'$ are conjugate for the holomorphic and anti-holomorphic sectors separately.
Then, the corresponding primary operators have the same conformal dimensions. We have that
\begin{equation}
{\rm Reps\ (1)\ and\  (2)\ conjugate}:\
t^{(1)}_a= t_a\ , \quad \tilde t^{(1)}_a= \tilde t_a \ ,\quad t^{(2)}_a= -t_a^*\ , \quad \tilde t^{(2)}_a= -\tilde t_a^*\ ,
\label{conjj}
\end{equation}
where the first equality is just a convenient renaming to avoid superscripts.
In this paper we will not compute the $\l$-deformed three-point function involving
only primary fields (for the $k_L=k_R$ case this was done in \cite{Georgiou:2016iom}).
Also, correlators with two currents and one affine primary field are zero for $\l=0$ and stay so in the deformed theory as well.

\no
In the following sections, we will compute the two- and three-point function of currents and for
primary fields  in the left--right asymmetric case, but now for $\l\neq 0$. From these we will extract the corresponding
anomalous dimensions, OPEs, equal time commutators and classical Poisson brackets.
Such correlators will be denoted by $\langle \cdots \rangle_\l$ in order to
distinguish them from those evaluated at the CFT point, that is for vanishing $\l$.

\section{Current correlators and anomalous dimensions}

\subsection{Two-point functions }

In this subsection, we will evaluate the correlator of two holomorphic or two antiholomorphic currents. From this we will read the anomalous dimensions of the currents to all-orders in
the deformation parameter $\lambda$.
At ${\cal O}(\l^n)$, the correlation function $\langle J_a(x_1) J_b(x_2)\rangle_\l$
involves the sum of the expressions
\be
\label{expansion}
\begin{split}
& \langle J_{a}(x_1) J_b(x_2)  \rangle_{\l}^{(n)}
 = {1\ov n!} \left(-\frac{\l}{\pi}\right)^n
\\
& \phantom{xx} \times \int \text{d}^2z_{1\dots n} \langle J_{a}(x_1) J_b(x_2) J_{a_1}(z_1)\cdots J_{a_n}(z_n)\rangle
\langle \bar J_{a_1}(\bar z_1)\cdots \bar J_{a_n}(\bar z_n) \rangle\ ,
\end{split}
\ee
where $\text{d}^2z_{1\dots n}=\text{d}^2z_1\cdots \text{d}^2z_n$.  The dependence on the levels $k_L$ and $k_R$ arises
from the evaluation of the correlation functions appearing in the integrand, at the CFT point.
The right hand side of the above expression will in general be a function of the
$x_i$'s as well as of the $\bar x_i$'s. For notational convenience we have omitted this
dependence on the left hand side which we will consistently follow throughout
this paper.

\no
On general grounds, the two-point function can be cast in the form
\be\label{normJ}
\langle J_a(x_1) J_b(x_2)\rangle_{\l} = {\d_{ab} \ov x_{12}^{2+{\g_L}}
\bar x_{12}^{\g_L}}\ ,
\ee
where the deformation is encoded in the anomalous dimension $\g_L$ of the holomorphic current.\footnote{We will keep characterizing
$J_a$ and $\bar J_a$ as holomorphic and anti-holomorphic, respectively, even though when the deformation is turned on they are
no longer.} We will denote the anomalous dimension of the anti-holomorphic currents by $\g_R$.
Also, \eqn{normJ} implies that the holomorphic and anti-holomorphic dimensions of $J_a$ and $\bar J_a$ are
$(1+\nicefrac{\g_L}{2},\nicefrac{\g_L}{2})$ and $(\nicefrac{\g_R}{2},1+\nicefrac{\g_R}{2})$, respectively.
The perturbative calculation up to order $\mathcal{O}(\l^3)$ results to the following expression
\be
\g_L(k_L,k_R,\l) = {c_G\ov k_L} \l^2 - 2 {c_G\ov \sqrt{k_L k_R}} \l^3 + \cdots\ , \
\label{glpert}
\ee
where the ellipses denote higher order terms in the $\l$ as well as in the
$1\ov k_{L,R}$ expansion.
This expression arises from an analogous computation in \cite{Georgiou:2016iom} for the left--right symmetric case just by
keeping track of the factors of $k_L$ and $k_R$, hence the omission of the details.

To proceed we need the exact in $\l$ beta-function. This is found using results of \cite{LeClair:2001yp}\footnote{
We use Eq. (3.4) of that reference where in order to conform with our notation we let
\begin{equation*}
k_{R,L}\mapsto 2k_{R,L}\,,\quad
g \mapsto \frac{2\lambda}{\sqrt{k_Lk_R}}\,,\quad
C_{\text{Adj}} \mapsto c_G\ .
\end{equation*}
The logarithm of the length scale in $\beta_g$ is replaced by $t=\ln\mu^2$ which effectively flips the overall sign.
}
\be
\boxed{\
{\text{d}\l\ov \text{d}t} = -{c_G\ov 2 \sqrt{k_Lk_R}} {\l^2 (\l-\l_0) (\l-\l_0^{-1}) \ov (1-\l^2)^2}\ ,
\qq \l_0 = \sqrt{k_L\ov k_R}\ }\ ,
\label{betal}
\ee
where $t=\ln \m^2 $ , with $\m$ being the energy scale.
There are
three fixed points, at $\l=0$, at $\l= \l_0$ and at $\l=\l_0^{-1}$ in which the beta-function vanishes.
The first at $\l=0$ is the usual UV stable fixed point, present in the left--right symmetric case, as well. However,
the other two fixed points are new. To investigate their nature we will assume through out the paper with no loss of generality
that $\l_0 < 1$.
Then, it is easy to see that, the fixed point at  $\l=\l_0$ is IR stable whereas that for $\l=\l_0^{-1}> 1$ is UV stable.
The first of these points is reached from $\l=0$ under an RG flow.
The second one can only be reached from large values of $\l$.
Flowing between the two fixed points involves passing thought the strong coupling region at $\l=1$.\footnote{By
 strong coupling region we mean the
region where the $\beta$-function develops poles.}
We will argue that the region with $\l>1$ should be dismissed and therefore the fixed point at $\l=\l_0^{-1}$ is
unphysical.
These are new feature of the left--right asymmetric models not present in the left--right symmetric case.
We will see that at these new fixed points the anomalous dimensions of the operators we will compute
below are generically non-vanishing.

\no
The above beta-function is well defined under the correlated limit in which
$\l\to \pm 1$ and in addition the levels become extremely large. Specifically,
\be
\l= \pm 1 - {b \ov (k_L k_R)^{1/6}}\ , \qq k_{L,R}\to \infty\ ,
\label{lpso}
\ee
where $b$ is the new coupling constants and where the limit is taken
for both signs independently. This is a direct analogue  of the pseudochiral model limit $\l\to -1$ \cite{Georgiou:2016iom}
which can be taken not only in the beta-function and anomalous dimensions, but also at the level of the all-loop effective action
of \cite{Sfetsos:2013wia} for $k_L=k_R$. It is interesting that this limit exists even though in the left--right asymmetric
case we do not know the corresponding all-loop effective action. Note that one cannot take the non-Abelian T-duality limit
$\l\to 1$ which exists for the left--right symmetric case only \cite{Sfetsos:2013wia}.\footnote{This limit is of the form \eqn{lpso} but with
 $ (k_L k_R)^6\mapsto k$ and also involves an expansion of the group element of the WZW model action.}
The distinction between the $\l\to 1$ and $\l\to -1$ limits ceases to exist in the left--right asymmetric case.

\no
As in \cite{Georgiou:2015nka} we may deduce by examining the
Callan--Symanzik equation the form of the anomalous dimension to all orders in $\l$.
The appropriate ansatz is of the form
\be
\g_L(k_L,k_R,\l)= {c_G\ov \sqrt{k_L k_R}} {\l^2\ov (1-\l^2)^3} f(\l ; \l_0) + \cdots \ ,
\label{anomgl}
\ee
where now the ellipses denote higher order terms in the large level expansion.
The function $f(\l ;\l_0)$ is a to be determined and should be analytic in the complex $\l$-plane. It may depend  on
the levels  $k_L$ and $k_R$ as parameters only via their ratio since we are interested in the leading order behaviour in the levels.
This explains the presence of the parameter $\l_0$. Note also that this ansatz remains finite
under the limit \eqn{lpso}.

\no
In \cite{Kutasov:1989aw} it was argued, using path integral arguments and manipulations, that the theory should be invariant under the transformation
\be
{\rm For}\ k_{L,R} \gg 1 : \qquad k_L\mapsto -k_R\ ,\quad  k_R\mapsto -k_L\ ,\quad \l\mapsto {1\ov \l} \ .
\label{symmas}
\ee
We emphasize that this statement is true without including the parity transformation $z\leftrightarrow \bar z$.
In implementing the above in various expressions one should be careful and treat it
as an analytic continuation when square roots appear. Hence, we better write
$(k_L,k_R)\mapsto e^{i\pi} (k_R,k_L)$ and $1-\l\mapsto e^{-i\pi} \l^{-1}(1-\l)$.
One easily sees that this is a symmetry of  the beta-function equation \eqn{betal}.
Imposing it to be a symmetry of the anomalous dimension \eqn{anomgl}, i.e.
\be
\g_L(-k_R,-k_L,\l^{-1})=\g_L(k_L,k_R,\l) \quad  \Longrightarrow\quad
\l^2 f(\l^{-1}; \l_0^{-1})= f (\l ; \l_0)\ ,
\ee
implying that $ f (\l ; \l_0)$ is a second order polynomial in $\l$ with coefficients depending on
$\l_0$ and related via the above symmetry.
The matching of its coefficients with the perturbative result \eqn{glpert}
gives $ f (\l ; \l_0)= \l_0 (\l-\l_0^{-1})^2 $. Hence we obtain for the holomorphic current' anomalous dimension the exact in $\l$ expression
\be
\boxed{\
\g_L(k_L,k_R,\l)= {c_G\ov  k_R} {\l^2(\l-\l_0^{-1})^2\ov (1-\l^2)^3} + \cdots \ }
 \ .
\label{anomglre}
\ee
Similar considerations lead to the anomalous dimension of the anti-holomorphic current with the result being
\be
\boxed{\
\bar\g_R(k_L,k_R,\l)= {c_G\ov  k_L} {\l^2(\l-\l_0)^2\ov (1-\l^2)^3} + \cdots \ }\ .
\label{anomgrre}
\ee
Notice that under the limit \eqn{lpso} both anomalous dimensions remain finite.

\no
Obviously, for $\l=0$ the anomalous dimensions vanish. However, this is no longer true for the other fixed points of the
beta-function. We have that
\be
\label{ffpp}
\begin{split}
&  \l= \l_0:\qq  \ \ \g_L(k_L,k_R,\l_0)={c_G\ov k_R-k_L}\ ,\qq  \g_R(k_L,k_R,\l_0) = 0  \ ,
\\
&  \l= \l_0^{-1}:\qq  \g_L(k_L,k_R,\l_0^{-1})= 0\ ,\qq\qq \  \g_R(k_L,k_R,\l_0^{-1})  = {c_G\ov k_L-k_R}\ .
\end{split}
\ee
The presence of non-vanishing anomalous dimensions at the new fixed points is indicative of the fact that the new CFT at these
points are such that the original currents have acquired new characteristics.
We will return to this point later.

\no
To further check the validity of \eqn{anomglre} we have performed a tedious
perturbative computation at ${\cal O}(\l^4)$ whose details
are presented in appendix \ref{fourloop}. We found perfect agreement with the prediction of \eqn{anomglre} .

\subsection{Current composite operators}

In this subsection we will compute the two-point correlator of  the composite operator that deforms the CFT.
From this correlator we  will extract the  exact in
the deformation parameter $\lambda$ dimension of this operator. In particular,
we will compute the dimension of the current-bilinear operator
\be
{\cal O}(z,\bar z)=J_a(z)\bar J_a(\bar z)\ ,
\ee
 which actually drives the deformation
as in \eqref{thir}. We will do this in two different ways which will lead to the same result and are conceptually complementary.

In the first method we will use the geometry in the space of couplings as presented in \cite{Kutasov:1989dt} and used in the
present context when $k_L=k_R$ in \cite{Georgiou:2015nka}.
We have to evaluate the two-point function $G=\langle{\cal O}(x_1,\bar x_1){\cal O}(x_2,\bar x_2)\rangle$ and
read off the Zamolodchikov metric $g=|x_{12}|^4\langle{\cal O}(x_1,\bar x_1){\cal O}(x_2,\bar x_2)\rangle$.
It takes the form
\begin{equation}
\label{dkjdfk}
G\sim |x_{12}|^{-4}\left(1+\gamma^{({\cal O})}\ln\frac{\varepsilon^2}{|x_{12}|^2}\right)\,,\quad
g=g_0-2s\nabla_\lambda\beta+{\cal O}(s^2)\,,
\end{equation}
where $s=\ln(|x_{12}|^2\mu^2)$ and the $\beta$-function is given by \eqref{betal}.
The finite part of the two-point function was calculated as in  \cite{Kutasov:1989dt} (see \cite{Georgiou:2015nka} for a detailed derivation) and reads
\begin{equation}
\label{gope}
g_0=\frac{\dim G}{(1-\lambda^2)^2}\,.
\end{equation}
Upon this we built the connection that appears in the covariant derivative with respect to $\lambda$.
In our case we have just one coupling constant $\l$ and it turns out that the anomalous dimension of the composite operator is
simply given by
\begin{equation}
\label{kjefkjefk1}
\gamma^{({\cal O})}=2\nabla_\lambda\beta=2\partial_\lambda\beta+\beta\frac{\partial_\lambda g_0}{g_0}\ .
\ee
Using \eqn{betal} and \eqn{gope}, we find specifically that
\be
\label{kjefkjefk}
\boxed{\
\gamma^{({\cal O})}(k_L,k_R,\l)
=c_G\lambda\,\frac{3(\l_0+\l_0^{-1})\lambda(1+\lambda^2)-2(1+4\lambda^2+\lambda^4)}{\sqrt{k_Rk_L}(1-\lambda^2)^3}
+ \cdots \ }\  .
\ee
This respects the symmetry \eqn{symmas} since $\gamma^{({\cal O})}(-k_R,-k_L,\l^{-1})= \gamma^{({\cal O})}(k_L,k_R,\l)$.
Moreover, in the IR fixed point we find that
\be
\gamma^{({\cal O})}(k_L,k_R,\l_0) = {c_G\ov k_R-k_L}\ ,
\ee
that is at the fixed point of the RG flow at $\l=\l_0$ the anomalous dimensions of the composite operator
equals the sum of anomalous dimensions of $J_a$ and $\bar J_a$, as it should be since the two CFTs are decoupled.
Unlike, the left--right symmetric case for which $\gamma^{({\cal O})}$ is strictly non-positive,
here it doesn't have a definite sign for the RG in $\l\in (0,\l_0)$.
It starts negative, then it develops a minimum and finally it reaches the positive value given above. This is expected since the
perturbation is relevant and irrelevant near $\l=0$ and near $\l=\l_0$, respectively.

It is instructive to also derive \eqref{kjefkjefk}, using our method, i.e. low order perturbative results combined with the symmetry \eqn{symmas}. The leading order one-loop  contribution is
\ba
&&\langle{\cal O}(x_1,\bar x_1){\cal O}(x_2,\bar x_2)\rangle^{(1)}_\l
=-\frac{\lambda}{\pi}\int\text{d}^2z\langle J_a(x_1)J_b(x_2)J_c(z)\rangle\,
\langle \bar J_a(\bar x_1)\bar J_b(\bar x_2)\bar J_c(\bar z)\rangle
\nonumber\\
&&\qq\qq \phantom{xx}
=\frac{c_G\dim G\,\lambda}{\sqrt{k_Lk_R}\pi|x_{12}|^2}
\int\frac{\text{d}^2z}{(x_1-z)(x_2-z)(\bar x_1-\bar z)(\bar x_2-\bar z)}\,,
\label{compl}
\\
&& \qq\qq \phantom{xx}
=-\frac{2c_G\dim G\,\lambda}{\sqrt{k_Lk_R}|x_{12}|^4}\,\ln\frac{\varepsilon^2}{|x_{12}|^2}\,.
\nonumber
\ea

\no
Turning next to the two-loop contribution we have that
\ba
&& \langle{\cal O}(x_1,\bar x_1){\cal O}(x_2,\bar x_2)\rangle^{(2)}_\l
=\frac{\lambda^2}{2!\pi^2}\int\text{d}^2z_{12}\langle J_a(x_1)J_b(x_2)J_c(z_1)J_d(z_2)\rangle\,
\nonumber\\
&&\qq\qq\phantom{xxxx}
\times \langle \bar J_a(\bar x_1)\bar J_b(\bar x_2)\bar J_c(\bar z_1)\bar J_d(\bar z_2)\rangle\,.
\ea
To proceed we evaluate the four-point function using the Ward identity \eqref{WardIdent}
\begin{equation}
\begin{split}
& \langle J_{a_1}( z_1) J_{a_2}( z_2) J_{a_3}( z_3) J_{a_4}( z_4)\rangle=
\frac{1}{k_L}\left(\frac{f_{a_1a_3e}f_{a_2a_4e}}{ z_{12} z_{13} z_{24} z_{34}}
-\frac{f_{a_1a_4e}f_{a_2a_3e}}{ z_{12} z_{14} z_{23} z_{34}}\right)
\\
&\qq\qq\qq
+\frac{\delta_{a_1a_2}\delta_{a_3a_4}}{ z_{12}^2 z_{34}^2}
+\frac{\delta_{a_1a_3}\delta_{a_2a_4}}{ z_{13}^2 z_{24}^2}
+\frac{\delta_{a_1a_4}\delta_{a_2a_3}}{ z_{14}^2 z_{23}^2}\,
\end{split}
\end{equation}
and after some effort and heavy use of the identity
\begin{equation*}
\frac{1}{(x_1-z)(z-x_2)}=\frac{1}{x_{12}}\left(\frac{1}{x_1-z}+\frac{1}{z-x_2}\right)\,,
\end{equation*}
we find at leading order in $k_{L,R}$ that
\begin{equation}
\langle{\cal O}(x_1,\bar x_1){\cal O}(x_2,\bar x_2)\rangle^{(2)}_\l
=\frac{3c_G\dim G\,\lambda^2}{|x_{12}|^4}\,\left(\frac{1}{k_R}+\frac{1}{k_L}\right)\ln\frac{\varepsilon^2}{|x_{12}|^2}\,.
\end{equation}
Either from the form of the Callan--Symanzik equation, or by demanding a well-defined behaviour in the limit \eqref{lpso}
and then by employing the symmetry \eqref{symmas} we find the all-loop expression
\begin{equation}
\label{kjflflmdkd}
\gamma^{({\cal O})}=c_G\lambda\,
\frac{3(\l_0+\l_0^{-1})\lambda(1+\lambda^2)-2(1+4a_2 \lambda^2+\lambda^4)}{\sqrt{k_Rk_L}(1-\lambda^2)^3} + \dots
\end{equation}
where the coefficient $a_2$ can not be determined by the symmetry arguments.
 One way to determine it is to further compute the ${\cal O}(\l^3)$ perturbative contribution.
However, it is much easier to just demand for consistency that $\gamma^{({\cal O})}$
equals the sum of the anomalous dimensions of the currents $J_a$ and $\bar J_a$ at the fixed point $\lambda=\lambda_0$.
This fixes $a_2=1$ and therefore \eqref{kjflflmdkd} matches \eqref{kjefkjefk}.

\subsection{Three-point functions}

Similarly to the two-point function case we use the
perturbative result for the three-point function of holomorphic currents given by
\begin{equation}
\label{jfkfkksk}
\langle J_a(x_1) J_b(x_2) J_c(x_3)\rangle_\l
=\left[
\frac{1}{\sqrt{k_L}}\left(1+\frac32\,\lambda^2\right) -{1\ov \sqrt{k_R}}\lambda^3\right]
\frac{f_{abc}}{x_{12}x_{13}x_{23}} +\cdots  \ ,
\end{equation}
which follows from the analogous computation in \cite{Georgiou:2016iom} in the left--right symmetric case by modifying appropriate
the various terms to take into account the different levels.
The ansatz for the all-loop expression takes the form
\begin{equation}
\label{kflflk}
\langle J_a(x_1) J_b(x_2) J_c(x_3)\rangle_\l=\frac{g(\l ; \l_0)}{\sqrt{k_L(1-\lambda^2)^3}} \frac{f_{abc}}{\,x_{12}x_{13}x_{23}}\ ,
\end{equation}
where as before $g(\lambda ; \l_0)$ is everywhere analytic
and should be determined by imposing the symmetry \eqn{symmas} and matching with the perturbative result \eqn{jfkfkksk}. Imposing first the symmetry we obtain that
\be
\l_0 \l^3 g(\l^{-1}; \l_0^{-1})= - g(\l; \l_0)\ ,
\ee
implying that $ g (\l ; \l_0)$ is a third order polynomial in $\l$ with coefficients depending on  $\l_0$.
Simple algebra, taking also into account the perturbative result \eqn{jfkfkksk}, gives that $g (\l ; \l_0)= 1- \l_0\l^3$.
Hence, the exact in $\l$ three-point function of holomorphic function reads
\begin{equation}
\label{kflflkres}
\boxed{\
\langle J_a(x_1) J_b(x_2) J_c(x_3)\rangle_\l=\frac{1-\l_0\l^3}{\sqrt{k_L(1-\lambda^2)^3}}
\frac{f_{abc}}{\,x_{12}x_{13}x_{23}} + \dots \ }\ .
\end{equation}
In a similar fashion the three-point function for the anti-holomorphic currents is given by
\begin{equation}
\label{kflflkrer}
\boxed{\
\langle \bar J_a(\bar x_1) \bar J_b(\bar x_2) \bar J_c(\bar x_3)\rangle_\l=\frac{1-\l_0^{-1}\l^3} {\sqrt{k_R(1-\lambda^2)^3}} \frac{f_{abc}}{\,\bar x_{12}\bar x_{13}\bar x_{23}} + \cdots \ }\ .
\end{equation}

\no
It remains to compute the three-point function for mixed correlators. Using the, appropriately modified,
perturbative result of \cite{Georgiou:2016iom}
\begin{equation}
\label{jjjmi1}
\langle J_a(x_1) J_b(x_2) \bar J_c(\bar x_3)\rangle_\l
=\left(
\frac{\l}{\sqrt{k_L}}\ -{\l^2\ov \sqrt{k_R}}\right)
\frac{f_{abc} \bar x_{12}}{x_{12}^2\bar x_{13}\bar x_{23}} +\cdots  \ ,
\end{equation}
and imposing the symmetry \eqn{symmas} to an appropriate ansatz for the all-loop result
we obtain that
\be
\label{jjjmi2}
\boxed{\
\langle J_a(x_1) J_b(x_2) \bar J_c(\bar x_3)\rangle_\l=
{\l (\l^{-1}_0-\l)\ov \sqrt{k_R (1-\l^2)^3}}
\frac{f_{abc} \bar x_{12}} {x_{12}^2 \bar x_{13}\bar x_{23}} + \dots \ }\  .
\end{equation}
Similarly, for the other mixed three-point correlators we have that
\be
\label{jjjmi3}
\boxed{\
\langle \bar J_a(\bar x_1)\bar  J_b(\bar x_2)  J_c( x_3)\rangle_\l=
{\l (\l_0-\l)\ov \sqrt{k_L (1-\l^2)^3}}
\frac{f_{abc}  x_{12}} {\bar x_{12}^2 x_{13} x_{23}} + \dots \ }\ .
\end{equation}

\subsection{The IR fixed point}

The anomalous dimension $\g_R=0$ at $\l=\l_0$ according to \eqn{ffpp},
so that $\bar J_a$ remains with dimension $(0,1)$. This implies that $\bar J_a$ can be regarded as an antiholomorhic current
not only at $\l=0$ but also at $\l=\l_0$.
In addition, at $\l=\l_0$ the prefactor on the r.h.s. of \eqn{kflflkrer} becomes $1/\sqrt{k_R-k_L}$. This suggests that
under an RG flow the new conformal point at $\l=\l_0$ is reached in the IR and at this point
the anti-holomorphic current $\bar J_a$ generates
the same current algebra but with a smaller level $k_R\mapsto k_R-k_L$.

\no
The nature of
the holomorphic current at $\l=\l_0$ is more delicate to determine. Its anomalous dimension $(1+\nicefrac{\g_L}{2},\nicefrac{\g_L}{2})$ with
$\g_L$  given in \eqn{ffpp} implies, after recalling that we are in the $k_{L,R}\gg 1$ regime,
that it corresponds to an operator transforming in the adjoint representation for a current algebra at level $k_R-k_L$.
Indeed, using the OPE's in \eqref{OPEcurrents} below evaluated at $\lambda=\lambda_0$, we easily see that
\ba
&&J_a(x_1)J_b(x_2)=\frac{\delta_{ab}}{x_{12}^{2}}+\frac{k_R+k_L}{\sqrt{k_Lk_R(k_R-k_L)}}\frac{f_{abc}J_c(x_2)}{x_{12}}
+\frac{1}{\sqrt{k_R-k_L}}\frac{f_{abc}\bar J_c(\bar x_2)\bar x_{12}}{x^2_{12}}+\cdots\,,
\nonumber
\\
&&\bar J_a(\bar x_1)\bar J_b(\bar x_2)=\frac{\delta_{ab}}{\bar x_{12}^{2}}+
\frac{1}{\sqrt{k_R-k_L}}\frac{f_{abc}\bar J_c(\bar x_2)}{\bar x_{12}}+\cdots\,,
\label{conpoi}
\\
&&J_a(x_1)\bar J_b(\bar x_2)=\frac{1}{\sqrt{k_R-k_L}}\frac{f_{abc} J_c(x_2)}{\bar x_{12}}+\cdots\ ,
\nonumber
\ea
where we emphasize that these OPE's are valid to leading order in the large level expansion (beyond the Abelian limit).
Before proceeding, note also the curious fact that these OPE's are invariant under $(k_L,k_R)\mapsto -(k_R,k_L)$ which is the
remnant of the symmetry \eqn{symmas} once we have fixed $\l=\l_0$.

The middle of the above OPE's is indeed a current algebra theory
$G_{k_R-k_L}$, as advertized. The third line shows that $J_a$ transforms non-trivially under $\bar \J_a$.
 From the form of this OPE and the
anomalous dimensions that $J_a$ has acquired it appears as if $J_a$ is a composite operator of the
form $J_a = \tilde \Phi_{b,a} \tilde J_b$, where $\tilde \Phi$ is
a field transforming in the adjoint representation similar to \eqn{adjjj} and $\tilde J_a$ generates a current
algebra  (the tilded symbol will be explained shortly). This interpretation should be considered with caution as far as the
holomorphic sector is concerned. In that sector the theory is not a current algebra theory, in fact as we will argue shortly
it is a coset CFT. As such it does not possess currents as holomorphic objects but their counterparts which are
the non-Abelian parafermions \cite{Bardakci:1990ad} with dimensions deviating from unity by $1/k$-corrections,
These have not been studied at the quantum level \cite{Bardakci:1992kr} as much as their Abelian counterparts \cite{Fateev:1991bv}.
Similarly, for $\tilde \Phi_{a,b}$ the left representation is approximately in the limit of large levels similar to the adjoint one. These
comments explain the use of tilded symbols.
Nevertheless, for the anti-holomorphic sector the above statement is exact and indeed
one may check, using \eqn{adjjj}, with $k_R\mapsto k_R-k_L$ and the fact that the OPE between $\bar J_a $ and $\tilde J_b$ is regular,
that the third of \eqn{conpoi} is indeed reproduced.
Finally, we have checked that the OPE in the first line of \eqn{conpoi}
is reproduced by appropriately normalizing the structure constants of the quasi-current algebra for $\tilde J_a$
and by using that $\del \Phi_{c,a}\Phi_{c,b}\sim f_{abc}J_c$ and similarly for the  anti-homomorphic derivative, as well.
We will not present this computation here, not only because it is lengthy, but also since
the purpose of the above is to reinforce related arguments, made in the past, on the
nature of the CFT in the IR to which we now turn.

\no
There has been already a suggestion in \cite{LeClair:2001yp} based also on work in \cite{chiraliquids} that the RG flow is such that
\be
G_{k_L}\times G_{k_R} \quad \overset{\text{IR}}{\Longrightarrow}
 \quad {G_{k_L}\times G_{k_R-k_L}\ov G_{k_R}} \times G_{k_R-k_L}\ .
\label{sugge}
\ee
This was suggested based on the fact that the sum of the central charges for the left and the right sectors is expected to lower in
accordance with Zamolodchikov's $c$-theorem \cite{Zamolodchikov:1986gt}
and arguments that the difference of them should stay constant. One can argue that this is the case as follows.
For $k_R>k_L$ the central charge for the anti-holomorphic Virasoro algebra is larger that for the holomorphic one (see
\eqn{centraal}).
Hence, the theory is chiral, there is a gauge anomaly and also
the energy momentum tensor cannot be coupled anomaly free to a two-dimensional worldsheet metric.
We remedy the situation by making the levels of the current algebra and Virasoro algebra central charges equal by adding chiral matter.
These degrees of freedom do not participate in the RG flow and therefore the difference $k_R-k_L$, as well as that of that of the
Virasoro central charges has to be invariant and
their  values in the IR be the same as in the UV.\footnote{
We note for completeness that in the left--right symmetric case the RG flow is driven to a strong coupling regime towards
$\l=1$. In this case a mass gap develops which is consistent with the fact that in that regime the description is better in terms of the
non-Abelian T-dual of the principal chiral model (PCM) for the group $G$ which should have a mass gap \cite{Faddeev:1985qu} being canonically equivalent \cite{Curtright:1994be,Lozano:1995jx}
to the original PCM.
}
\no
The suggestion in \eqn{sugge} clearly satisfies these requirements and was presented as a unique solution for the end point of the RG flow
in the IR. This is further strongly reinforced by our findings as explained above. Nevertheless, these arguments are only suggestive and
a much better justification would be
to actually compute Zamolodchikov's $c$-function and its dependence on $\l$. This is left for future work.

\no
Finally, note that, due to \eqn{symmas} the theory for $\l>1$ and positive integers as levels of the two current algebras
is equivalently to a theory with $\l<1$ but now with negative levels of the current algebras.
Hence, unitarity is violated and this makes this region unphysical. In accordance,
the fixed point at $\l=\l_0^{-1}$ is excluded from our discussion.

\section{Primary field correlators and anomalous dimensions}
In this section, we calculate the all-loop anomalous dimensions of affine primary operators, as well as the three-point functions
of one current (holomorphic or antiholomorhic) with two affine primary operators.

The perturbative computation up to ${\cal O}(\l^3)$ and to ${\cal O}(1/k_{L,R})$, for
the two-point function of primary fields can be found by appropriately modifying the corresponding computation in
\cite{Georgiou:2016iom} done in the left--right symmetric case. The result is
\ba
&& \langle \Phi^{(1)}_{i,i'}(x_1,\bar x_1) \Phi^{(2)}_{j,j'}(x_2, \bar x_2)\rangle_\l =
{1\ov x_{12}^{2 \D_R} \bar x_{12}^{2 \bar \D_{R'}}}
\bigg[\left(1+\l^2 \left({c_R\ov k_L} + {c_{R'}\ov k_R}\right) \ln{\varepsilon^2\ov |x_{12}|^2}\right) (\mathbb{I}_R \otimes \mathbb{I}_{R'})_{ii',jj'}
\nonumber\\
&& \qq\qq\qq -2\l {1+\l^2\ov\sqrt{ k_L k_R}}   \ln{\varepsilon^2\ov |x_{12}|^2} 
{(t_a \otimes \tilde t_a^*)_{ii',jj'}}\bigg] + {1\ov k_{L,R}}{\cal O}(\l^4)\ .
\label{corrlf1}
\ea
Proceeding as in \cite{Georgiou:2016iom} there is a $\lambda$-independent matrix $U$ chosen such that
\begin{equation}
{(t_a \otimes \tilde t_a^*)_{IJ}}= U_{IK} N_{KL} (U^{-1})_{LJ}\ ,\qquad N_{IJ}=N_I \delta_{IJ}\ ,
\end{equation}
where the $N_I$'s are the eigenvalues of the matrix $t_a \otimes t_a^*$ and where
we have adopted the double index notation $I=(ii')$.
In the rotated basis
\begin{equation}
\widetilde \Phi^{(1)}_I = (U^{-1})_I{}^J \Phi^{(1)}_J \ ,\qq  \widetilde \Phi^{(2)}_I = U_I{}^J \Phi^{(2)}_J\ ,
\end{equation}
the correlator \eqn{corrlf1} becomes diagonal
\be
\langle \widetilde \Phi^{(1)}_I(x_1,\bar x_1) \widetilde \Phi^{(2)}_J(x_2, \bar x_2)\rangle_\l =
{\delta_{IJ} \ov x_{12}^{2 \D_R} \bar x_{12}^{2 \bar \D_{R'}}}
\left(1 + \d^{(\Phi)}_I \ln{\varepsilon^2\ov |x_{12}|^2} \right) \ ,
\ee
where
\be
\d^{(\Phi)}_I = - 2 \l {1+\l^2\ov \sqrt{k_L k_R}} N_I + \l^2\left({c_R\ov k_L} + {c_{R'}\ov k_R}\right) + {{\cal O}(\l^4)}\ .
\label{corrl2}
\ee
To determine the exact anomalous dimension of the general primary field we first realize that we should
include in the above expression the level-dependent part coming
from the CFT dimensions of $\D_R$ and $\bar \D_{R'}$ in \eqn{ddcft} up to order $1/k_{L,R}$.
Hence the anomalous dimension is given by
\begin{equation}
\begin{split}
&
\g^{(L)}_{I; R,R'}(k,\l)\big |_{\rm pert} = {c_R\ov  k_L} +  \d^{(\Phi)}_I=
\\
& \quad = {c_R\ov  k_L}(1+\l^2) + {c_{R'}\ov  k_R}\l^2  - 2{\l(1+\l^2)\ov \sqrt{k_L k_R}} N_I  +{{\cal O}(\l^4)}\ .
\end{split}
\label{petff}
\end{equation}
Using the symmetry \eqn{symmas} and the corresponding transformation for the primary fields and
representation matrices found in \cite{Georgiou:2016iom}
\be
\label{ffsym}
\Phi^{(1)}_{i,i'}  \leftrightarrow   \Phi^{(2)}_{i',i}\ ,\qq
(t^{(1)}, t^{(2)}) \leftrightarrow (\tilde t^{(2)}, \tilde t^{(1)})\ ,
\ee
we have for the all-loop anomalous dimensions the relation
 \begin{equation}
\g^{(L)}_{I;R,R'}(-k_R,-k_L,\l^{-1}) =  \g^{(L)}_{I;R',R}(k_L,k_R,\l)\ .
\end{equation}
Following our standard procedure with an appropriate ansatz we find that
\begin{equation}
\boxed{\
\g^{(L)}_{I; R,R'}(k_L,k_R,\l) =
{1\ov 1-\l^2}\left( {c_R\ov k_L} + { c_{R'}\ov k_R} \l^2 - 2 {\l\ov \sqrt{k_L k_R}}  N_I \right)\ }\,,
\end{equation}
and similarly  that
\begin{equation}
\boxed{\
\g^{(R)}_{I; R,R'}(k_L,k_R,\l) =
{1\ov 1-\l^2}\left( {c_{R'}\ov k_R} + { c_{R}\ov k_L} \l^2 - 2 {\l\ov \sqrt{k_L k_R}}  N_I\right)\ }\  .
\end{equation}
At the fixed point at $\l=\l_0$ none of the anomalous dimensions
vanishes identical for all possible representations. Depending on the representations $R$, $R'$ and the multiplicity number $N_I$,
there seem to be values of $\l$ for which the anomalous dimensions of specific fields may vanish.
Finally, the two-point function for conjugate primary fields take the form
\begin{equation}
\boxed{\
\langle \widetilde\Phi^{(1)}_{I}(x_1,\bar x_1)  \widetilde\Phi^{(2)}_{J}(x_2,\bar x_2)\rangle =  {\delta_{IJ}\ov
x_{12}^{\g^{(L)}_{I;R,R'}} \bar x_{12}^{\g^{(R)}_{I;R,R'}} }\
}\ .
\end{equation}

\no
For completeness, the mixed correlators involving two primary fields and one current can be evaluated
in a similar fashion resulting at
\begin{equation}
\langle J_a(x_3)\Phi_{i,i'}^{(1)}(x_1,\bar x_1)  \Phi_{j,j'}^{(2)}(x_2,\bar x_2)\rangle_\l=
- {  (t_a\otimes\mathbb{I}_{R'})_{ii',jj'}- \l_0 \lambda (\mathbb{I}_{R}\otimes\tilde t^*_a)_{ii',jj'}
\ov  \sqrt{k_L (1-\lambda^2)}  x_{12}^{2\Delta_R-1}\bar x_{12}^{2\bar\Delta_{R'}} x_{13} x_{23}} \ .
\end{equation}
Similar reasoning leads to
\be
\langle \bar J_a(\bar x_3)\Phi_{i,i'}^{(1)}(x_1,\bar x_1)  \Phi_{j,j'}^{(2)}(x_2,\bar x_2)\rangle_\l
=
\frac{(\mathbb{I}_{R}\otimes\tilde t^*_a)_{ii',jj'}-\l^{-1}_0\lambda (t_a\otimes\mathbb{I}_{R'})_{ii',jj'}}
{\sqrt{k_R(1-\lambda^2)}\,x_{12}^{2\Delta_R}\bar x_{12}^{2\bar\Delta_{R'}-1}  \bar x_{13} \bar x_{23}}\ .
\ee

\section{OPEs and Poisson structure}
\label{OPEPB}

Employing the two- and three-point functions of the currents and affine primaries found in the previous sections one may derive the OPE
at leading order in the $1/\sqrt{k_{L,R}}$ expansion and exact in the deformation parameter $\lambda$
\begin{equation}
\label{OPEcurrents}
\begin{split}
&J_a(x_1)J_b(x_2)=\frac{\delta_{ab}}{x_{12}^{2}}+a_L\frac{f_{abc}J_c(x_2)}{x_{12}}
+b_L\frac{f_{abc}\bar J_c(\bar x_2)\bar x_{12}}{x^2_{12}}+\dots\,,\\
&\bar J_a(\bar x_1)\bar J_b(\bar x_2)=\frac{\delta_{ab}}{\bar x_{12}^{2}}+a_R\frac{f_{abc}\bar J_c(\bar x_2)}{\bar x_{12}}
+b_R\frac{f_{abc} J_c( x_2) x_{12}}{\bar x^2_{12}}+\dots\,,\\
&J_a(x_1)\bar J_b(\bar x_2)=b_R\frac{f_{abc}\bar J_c(\bar x_2)}{x_{12}}
+b_L\frac{f_{abc} J_c(x_2)}{\bar x_{12}}+\dots\,,\\
&J_a(x_1) \Phi_{i,i'}^{(1)}(x_2,\bar x_2)=-
\frac{(t_a)_i{}^m\Phi^{(1)}_{m,i'}(x_2,\bar x_2)-\lambda_0\lambda(\tilde t_a^*)_{i'}{}^{m'}\Phi^{(1)}_{i,m'}(x_2,\bar x_2)}
{x_{12}\sqrt{k_L(1-\lambda^2)}}+\dots\,,\\
&\bar J_a(\bar x_1) \Phi_{i,i'}^{(1)}(x_2,\bar x_2)=
\frac{(\tilde t_a^*)_{i'}{}^{m'}\Phi^{(1)}_{i,m'}(x_2,\bar x_2)-\lambda_0^{-1}\lambda (t_a)_i^{m}\Phi^{(1)}_{m,i'}(x_2,\bar x_2)}
{\bar x_{12}\sqrt{k_R(1-\lambda^2)}}+\dots\,,\\
&\Phi_{I}^{(1)}(x_1,\bar x_1)\Phi_{J}^{(2)}(x_2,\bar x_2)=
 C_{IJK}\, \Phi_{K}^{(3)}(x_2,\bar x_2)+\dots\,,
\end{split}
\end{equation}
where $ C_{IJK}$ are the structure (numerical) constants of the affine primaries ring, and the various constants are given by
\begin{equation*}
\begin{split}
&a_L=\frac{1-\l_0\l^3}{\sqrt{k_L(1-\lambda^2)^3}}\ ,\qquad b_L={\l (\l_0^{-1}-\l)\ov \sqrt{k_R (1-\l^2)^3}}\,,
\\
&a_R=\frac{1-\l_0^{-1}\l^3} {\sqrt{k_R(1-\lambda^2)^3}}\ , \qquad b_R={\l (\l_0-\l)\ov \sqrt{k_L (1-\l^2)^3}}\,.
\end{split}
\end{equation*}
Using the above relations, we can evaluate the equal-time commutators of the currents and affine primaries via the time-ordered limiting procedure
\begin{equation*}
[F(\sigma_1,\tau),G(\sigma_2,\tau)]=\lim_{\varepsilon\to0}
\left(F(\sigma_1,\tau + \varepsilon)G(\sigma_2,\tau)-G(\sigma_2,\tau+\varepsilon)F(\sigma_1,\tau)\right)
\end{equation*}
and the limit representations of Dirac delta-function distribution
\begin{equation*}
\begin{split}
&\lim_{\varepsilon\to0}\left(\frac{1}{\sigma-i\varepsilon}
-\frac{1}{\sigma+i\varepsilon}\right)=2\pi i\,\delta_\sigma\,,\\
&\lim_{\varepsilon\to0}\left(\frac{\sigma+i\varepsilon}{(\sigma-i\varepsilon)^2}-
\frac{\sigma-i\varepsilon}{(\sigma+i\varepsilon)^2}\right)=2\pi i\,\delta_\sigma\,,\\
&\lim_{\varepsilon\to0}\left(\frac{1}{(\sigma-i\varepsilon)^2}
-\frac{1}{(\sigma+i\varepsilon)^2}\right)=-2\pi i\,\delta'_\sigma\,,
\end{split}
\end{equation*}
where $\delta_\sigma:=\delta(\sigma)$.

Using the above ${\cal O}(1/\sqrt{k_{L,R}})$, we find for the currents
\begin{equation}
\begin{split}
\label{currentcommut}
&[J_a(\sigma_1),J_b(\sigma_2)]=2\pi\, i\,\delta_{ab}\delta_{12}'+
2\pi\,f_{abc}\left(\,a_L J_c(\sigma_2)-b_L \bar J_c(\sigma_2)\right)
\delta_{12}\,,\\
&[\bar J_a(\sigma_1),\bar J_b(\sigma_2)]=-2\pi\, i\,\delta_{ab}\delta_{12}'+
2\pi\,f_{abc}\left(a_R \bar J_c(\sigma_2)-b_R J_c(\sigma_2)\right)
\delta_{12}\,,\\
&[J_a(\sigma_1),\bar J_b(\sigma_2)]=
2\pi\,f_{abc}\left(b_L J_c(\sigma_2)+b_R \bar J_c(\sigma_2)\right)\delta_{12}\  ,
\end{split}
\end{equation}
and for the primaries
\begin{eqnarray}
&&[\Phi^{(1)}_{i,i'}(\sigma_1),\Phi^{(2)}_{j,j'}(\sigma_2)]=0\,,\\
&&[J_a(\sigma_1),\Phi^{(1)}_{i,i'}(\sigma_2)]=-\frac{2\pi}{\sqrt{k_L(1-\lambda^2)}}\,
\left((t_a)_i{}^m\Phi^{(1)}_{m,i'}(\sigma_2)-\lambda_0\lambda(\tilde t_a^*)_{i'}{}^{m'}\Phi^{(1)}_{i,m'}(\sigma_2)\right)\,\delta_{12}\,,\nonumber\\
&&[\bar J_a(\sigma_1),\Phi^{(1)}_{i,i'}(\sigma_2)]=\frac{2\pi}{\sqrt{k_R(1-\lambda^2)}}\,
\left((\tilde t_a^*)_{i'}{}^{m'}\Phi^{(1)}_{i,m'}(\sigma_2)-\lambda_0^{-1}\lambda (t_a)_i{}^{m}\Phi^{(1)}_{m,i'}(\sigma_2)\right)\,\delta_{12}\,.\nonumber
\end{eqnarray}

\no
Next, we take the classical limit of \eqref{currentcommut}. The result
is the two-parameter deformation of the Poisson brackets for the isotropic PCM \cite{Balog:1993es} (in our conventions $f_{abc}$
are imaginary)
\begin{mathbox}
    \begin{align}
\label{Hungarians}
&\{I_\pm^a(\sigma_1),I_\pm^b(\sigma_2)\}_{\text P.B.}=-i\,\text{e}^2f_{abc}
\left((1\pm\rho)I_\mp^c(\sigma_2)-(1\mp\rho+2x(1\pm\rho))I_\pm^c(\sigma_2)\right)\delta_{12}\pm2\text{e}^2\delta_{ab}\,\delta_{12}'\, ,\nonumber
\\
&\{I_\pm^a(\sigma_1),I_\mp^b(\sigma_2)\}_{\text P.B.}=i\,\text{e}^2f_{abc}
\left((1+\rho)I_+^c(\sigma_2)+(1-\rho)I_-^c(\sigma_2)\right)\delta_{12}\,,
  \end{align}
\end{mathbox}

where we have rescaled the currents as
\begin{equation*}
J_a\mapsto -{1\ov \text{e}} I_+^a\,,\qq  \bar J_a\mapsto - {1\ov \text{e}}I_-^a\ ,
\end{equation*}
and the various parameters are
\be
\begin{split}
& \text{e}^2=4b^2_R(1-\rho)^{-2}=4b^2_L(1+\rho)^{-2} = {(\l_0^{1/2} + \l_0^{-1/2})^2\ov \sqrt{k_L k_R}} {\l^2  \ov (1-\l)(1+\l)^3}\ ,
\\
&
x=\frac{1+\lambda^2}{2\lambda}\,,\quad \rho=\frac{(1-\lambda_0)(1+\lambda)}{(1+\lambda_0)(1-\lambda) }\ .
\end{split}
\end{equation}
These parameters are invariant under the transformation \eqn{symmas} and so is the above algebra.
These brackets generalize Rajeev's extension \cite{Rajeev:1988hq} of the Poisson structure of the isotropic PCM.

\no
It is interesting to note that these Poisson brackets are isomorphic to
two commuting current algebras with levels $k_{L,R}$
\begin{equation}
\label{KacMoodyksks}
\boxed{
\{\Sigma^a_\pm(\sigma_1),\Sigma^b_\pm(\sigma_2)\}_{\text{P.B.}}=
-i\,f_{abc}\,\Sigma_\pm^c(\sigma_2)\delta_{12}\pm\frac{k_{L,R}}{2}\delta_{ab}\delta'_{12}
}\ ,
\end{equation}
where
\begin{equation*}
\Sigma^a_\pm={k_{L,R}\ov 4}\big( (1-\l)(1\pm \r) + 2\l\big)\Big(I_\mp^a -{1\ov \l} I^a_\pm\Big)\ .
\label{basch}
\end{equation*}
Note that in this decoupling form the symmetry \eqn{symmas} simply interchanges the two Poisson brackets in the algebra
\eqn{KacMoodyksks} since
using the above basis change we see that $\S_{\pm}\mapsto \S_{\mp}$.
Although the parameter $\lambda$ does not appear in the algebra \eqref{KacMoodyksks},
we expect that the effective Hamiltonian expressed in terms of $\Sigma_\pm^a$ depends on $\lambda$, as in the isotropic case.
This decoupled form of the algebra \eqn{KacMoodyksks} has also been observed in \cite{Delduc:2014uaa}.

\section{Conclusions}

In this work we investigate $\l$-deformed current theories based on a general semi-simple group but with a
left--right asymmetry induced by the different levels in the left and right sectors of the theory.
These left--right asymmetric theories are very interesting for several reasons.
They possess a new non-trivial fixed point compared to the left--right symmetric case and there is a smooth RG flow form the
the undeformed current algebra theory in the UV to a new CFT in the IR (see, \eqn{sugge}),
on the nature of which we gave strong arguments.
To do so by computing the all-loop anomalous dimensions of the left and right currents,
that of primary field operators in the aforementioned theories, as well as the exact in $\lambda$
three-point functions involving currents and/or primary fields.
Our computational method introduced in \cite{Georgiou:2015nka} and further developed in \cite{Georgiou:2016iom}
combines low order perturbative results with symmetry arguments.
The expressions for the aforementioned correlators allowed us to deduce the exact, in the deformation parameter, OPEs of the operators
involved and from these  OPEs their equal-time commutators.
We have found that the associated currents' Poisson brackets are a two-parameter deformation
of the canonical structure of the isotropic PCM which was found by purely classical algebraic methods \cite{Balog:1993es}.
Upon a suitable change of basis,
this Poisson bracket structure is isomorphic to two commuting current algebras with levels $k_{L,R}$.
In the isotropic case $k_L=k_R$, this Poisson structure was found to coincide with
Rajeev's one-parameter family (with the parameter $\r=0$ in \eqn{Hungarians}) of the deformed Poisson brackets of the isotropic PCM \cite{Georgiou:2016iom} and is
the canonical structure of the integrable $\lambda$-deformed $\sigma$-model \cite{Sfetsos:2013wia}.
It would be interesting to discover an effective action of the anisotropic non-Abelian Thirring model action as well.

\section*{Acknowledgements}

K. Siampos' work was partially supported by the \textsl{Germaine de Sta\"el} Franco--Swiss bilateral
 program 2015 (project no 32753SG). K. Siampos and K. Sfetsos would like to thank the
 TH-Unit at CERN for hospitality and financial support during the final stages of this project.

\appendix

\section{Useful integrals}
\label{parartima}

In this appendix we collect all the integrals which are needed in the calculations
and are evaluated by the use of Stokes'
theorem in the complex plane
\be
\int_D\text{d}^2z\left(\partial_z U+\partial_{\bar z} V\right)=
\frac{i}{2}\oint_{\partial D}\left(U\text{d}\bar z-V\text{d}z\right)\,.
\ee

In our regularization scheme we keep a small distance regulator $\varepsilon$ when two points coincide (for
further details on our regularization scheme the interested reader can consult \cite{Georgiou:2015nka,Georgiou:2016iom})
\begin{equation}
\label{id1}
\int \frac{\text{d}^2z}{(x_1-z)(\bar{z}-\bar{x}_2)}=\pi\ln|x_{12}|^2\,,\quad
\int \frac{\text{d}^2z}{(x_1-z)(\bar{z}-\bar{x}_1)}=\pi\ln\varepsilon^2\,,
\end{equation}

\be
\label{id2}
\int \frac{\text{d}^2z}{(x_1-z)^2(\bar{z}-\bar{x}_2)}=-\frac{\pi}{x_{12}}\ ,
\qq
\int \frac{\text{d}^2z}{( x_1- z)(\bar{z}-\bar{x}_2)^2}=-\frac{\pi}{\bar x_{12}}\,,
\end{equation}

\begin{equation}\label{id3}
\int \frac{\text{d}^2z}{(x_1-z)^2(\bar{z}-\bar{x}_2)^2}=0\,,
\end{equation}

\begin{equation}
\int\frac{\text{d}^2z}{(z-x_1)(\bar z-\bar x_2)^2}\,\ln |z-x_1|^2=
  \frac{\pi}{\bar x_{12}}\ln |x_{12}|^2\,,
\label{ln1b22}
\end{equation}

\begin{equation}
\int\frac{\text{d}^2z}{(z-x_1)^2(\bar z-\bar x_2)^2}\,\ln |z-x_1|^2=
 \frac{\pi}{|x_{12}|^2}\,,
\label{ln1b2}
\end{equation}

\be\label{ln2b2dif}
\int\frac{\text{d}^2z}{(z-x_2)^2(\bar z-\bar x_3)^2}\,\ln |z-x_1|^2=
\frac{\pi}{ \bar x_{13}}\left(\frac{1}{x_{23}}-\frac{\bar x_{12}}{x_{12}\bar x_{23}}\right)\,,
\ee

\begin{equation}
\label{12-same}
 \int {\text{d}^2 z\ov (z-x_1)(\bar z-\bar x_1)^2}=0\,.
\ee

\section{Current anomalous dimension at four-loop}
\label{fourloop}

In this appendix we would like to sketch the four-loop computation of the current operator anomalous dimension
at leading order in the large $k_{L,R}$ expansion; denoted as $\gamma^{(4)}_L$.
To proceed, we evaluate the four-loop contribution to the two-point function
\begin{equation}
\begin{split}
\label{hrfnlijrfm}
\langle J_a(x_1) J_b(x_2)\rangle^{(4)}_\l=
\frac{\lambda^4}{4!\pi^4}\,\int&\text{d}^2z_{1234}\,\langle J_a(x_1)J_b(x_2)J_{a_1}(z_1)J_{a_2}(z_2)J_{a_3}(z_3)J_{a_4}(z_4)\rangle
\\
&\times \langle \bar J_{a_1}(\bar z_1)\bar J_{a_2}(\bar z_2)\bar J_{a_3}(\bar z_3)\bar J_{a_4}(\bar z_4)\rangle\,.
\end{split}
\end{equation}
Employing the Ward identity \eqref{WardIdent} it is clear that the  ${\cal O}(\l^4)$
the current anomalous dimension is of the form
\begin{equation}
\label{klfldldhhuiei}
\gamma^{(4)}_L=c_G\lambda^4\,\left(\frac{\alpha_1}{k_L}+\frac{\alpha_2}{k_R}\right)\,, \quad
\alpha_1+\alpha_2=4\, ,
\end{equation}
where the last conditon is such that we match the four-loop contribution the left--right symmetric case when $k_L=k_R=k$
which was computed in  \cite{Georgiou:2016iom}. Hence, it is enough to find the contribution of the $1/k_R$-term.
Hence,
\begin{eqnarray}
&& \langle J_a(x_1) J_b(x_2)\rangle^{(4)}_\l \Big|_{1/k_R-{\rm term}}
=\frac{\lambda^4}{4!\pi^4k_R}\,\int\text{d}^2z_{1234}
\left(\frac{f_{a_1a_3e}f_{a_2a_4e}}{\bar z_{12}\bar z_{13}\bar z_{24}\bar z_{34}}
-\frac{f_{a_1a_4e}f_{a_2a_3e}}{\bar z_{12}\bar z_{14}\bar z_{23}\bar z_{34}}\right)
\\
&&\ \times \left(\frac{\delta_{aa_1}}{(x_1-z_1)^2}\langle J_b(x_2)J_{a_2}(z_2)J_{a_3}(z_3)J_{a_4}(z_4)\rangle
+\frac{\delta_{aa_2}}{(x_1-z_2)^2}\langle J_b(x_2)J_{a_2}(z_1)J_{a_3}(z_3)J_{a_4}(z_4)\rangle\right.\nonumber\\
&&\ +\left.\frac{\delta_{aa_3}}{(x_1-z_3)^2}\langle J_b(x_2)J_{a_1}(z_1)J_{a_2}(z_2)J_{a_4}(z_4)\rangle
+\frac{\delta_{aa_4}}{(x_1-z_4)^2}\langle J_b(x_2)J_{a_1}(z_1)J_{a_2}(z_2)J_{a_3}(z_3)\rangle\right)\nonumber\,,
\end{eqnarray}
where we have disregarded a bubble term.
The latter expression can be rewritten symbolically as
\begin{equation}
\label{jkflldldjkjs}
 \langle J_a(x_1) J_b(x_2)\rangle^{(4)}_\l \Big|_{1/k_R-{\rm term}} =\frac{\lambda^4}{4!\pi^4k_R}\,\int\text{d}^2z_{1234}
\left(I-II\right)\times\left(B+C+D+E\right)\,,
\end{equation}
where we consider only the Abelian part in the holomorphic four-point function.
To evaluate \eqref{jkflldldjkjs} it is a lengthy but straightforward computation which lays upon
using the point splitting formula
\begin{equation}
\label{pointsplit}
\frac{1}{(x_1-z)(z-x_2)}=\frac{1}{x_{12}}\left(\frac{1}{x_1-z}+\frac{1}{z-x_2}\right)\,,
\end{equation}
and the formulae of Appendix \ref{parartima}. In addition, in accordance to our regularization scheme we follow a specific order
in performing the integrations which we never violate, that is first the $z_4$ integration, then the $z_3$ one and so on and so forth.

\no
Among all possible terms, let us consider one of the terms in \eqref{jkflldldjkjs}
\begin{equation}
\label{wuisjsk}
\begin{split}
&\int\text{d}^2z_{1234}\,I\times B=\frac{1}{k_R}\int
\frac{\text{d}^2z_{1234}}{(x_1-z_1)^2}\langle J_b(x_2)J_{a_2}(z_2)J_{a_3}(z_3)J_{a_4}(z_4)\rangle
\frac{f_{aa_3e}f_{a_2a_4e}}{\bar z_{12}\bar z_{13}\bar z_{24}\bar z_{34}}\\
&=-\frac{c_G\delta_{ab}}{k_R}\int\frac{\text{d}^2z_{1234}}{(z_1-x_1)^2\bar z_{12}\bar z_{13}\bar z_{24}\bar z_{34}}\left(
\frac{1}{(z_2-x_2)^2z_{34}^2}-\frac{1}{(z_4-x_2)^2z_{23}^2}\right)\\
&\qq \qq \Longrightarrow\quad \int\text{d}^2z_{1234}\,I\times B=-\frac{c_G\delta_{ab}}{k_R}\left[\text{(i)}-\text{(ii)}\right]\,,
\end{split}
\end{equation}
with the apparent symbolic expression for the terms $(i)$ and $(ii)$.
To evaluate the term denoted by $(i)$ in \eqref{wuisjsk}, we integrate over $z_4$ and
employ \eqref{pointsplit}, \eqref{id2}, \eqref{12-same}
\begin{equation}
\text{(i)}=\pi\int\frac{\text{d}^2z_{123}}{(z_1-x_1)^2(z_2-x_2)^2\bar z_{12}\bar z_{13}\bar z_{23} z_{23}}\, .
\end{equation}
Then we integrate over $z_3$ and use \eqref{pointsplit}, \eqref{id1}
\begin{equation}
\text{(i)}=\pi^2\int\text{d}^2z_{12}\,\frac{\ln\frac{|z_{12}|^2}{\varepsilon^2}}{(z_1-x_1)^2(z_2-x_2)^2\bar z^2_{12}}\, ,
\end{equation}
next we integrate over $z_2$ and apply \eqref{id3}, \eqref{ln1b2}
\begin{equation}
\text{(i)}=\pi^3\int\text{d}^2z_{1}\,\frac{1}{(z_1-x_1)^2|z_1-x_2|^2}\, .
\end{equation}
Finally we integrate over $z_1$ and employ \eqref{pointsplit}, \eqref{id1}, \eqref{id2}
\begin{equation}
\label{iterm}
\text{(i)}=-\frac{\pi^4}{x_{12}^2}\left(1+\ln\frac{\varepsilon^2}{|x_{12}|^2}\right)\,.
\end{equation}
Similarly we evaluate the term $(ii)$ in \eqref{wuisjsk} to find that
\begin{equation}
\label{iiterm}
\text{(ii)}=\frac{2\pi^4}{x_{12}^2}\left(1+\ln\frac{\varepsilon^2}{|x_{12}|^2}\right)\,.
\end{equation}
Plugging \eqref{iterm} and \eqref{iiterm} into \eqref{wuisjsk} we find
\begin{equation}
\int\text{d}^2z_{1234}\,I\times B=\frac{c_G\delta_{ab}}{k_R}\frac{3\pi^4}{x_{12}^2}\left(1+\ln\frac{\varepsilon^2}{|x_{12}|^2}\right)\,.
\end{equation}
Working in a similar manner leads to
\begin{equation}
\label{kkdksiojs}
\begin{split}
&\int\text{d}^2z_{1234}\,I\times B=-\int\text{d}^2z_{1234}\,II\times B=
\frac{c_G\delta_{ab}}{k_R}\frac{3\pi^4}{x_{12}^2}\left(1+\ln\frac{\varepsilon^2}{|x_{12}|^2}\right)\,,\\
&\int\text{d}^2z_{1234}\,I\times C=-\int\text{d}^2z_{1234}\,II\times C
=\frac{c_G\delta_{ab}}{k_R}\frac{\pi^4}{x_{12}^2}\left(1+3\ln\frac{\varepsilon^2}{|x_{12}|^2}\right)\,,\\
&\int\text{d}^2z_{1234}\,I\times D=-\int\text{d}^2z_{1234}\,II\times E
=\frac{c_G\delta_{ab}}{k_R}\frac{3\pi^4}{x_{12}^2}\ln\frac{\varepsilon^2}{|x_{12}|^2}\,,\\
&\int\text{d}^2z_{1234}\,I\times E=-\int\text{d}^2z_{1234}\,II\times D
=\frac{c_G\delta_{ab}}{k_R}\frac{\pi^4}{x_{12}^2}\left(2+3\ln\frac{\varepsilon^2}{|x_{12}|^2}\right)\,,
\end{split}
\end{equation}
Plugging \eqref{kkdksiojs} into \eqref{jkflldldjkjs} we find that
\begin{equation}
\langle J_a(x_1) J_b(x_2)\rangle^{(4)}_\l \Big |_{1/k_R-{\rm term}}=\frac{c_G\delta_{ab}}{k_R\,x_{12}^2}\left(\frac12+\ln\frac{\varepsilon^2}{|x_{12}|^2}\right)\,,
\end{equation}
so $\alpha_2=1$ and therefore $\alpha_1=3$ in \eqref{klfldldhhuiei}.
Putting all these together we find that
\begin{equation}
\gamma^{(4)}_L=c_G\lambda^4\,\left(\frac{3}{k_L}+\frac{1}{k_R}\right)
\, ,
\end{equation}
which is consistent with the ${\cal O}(\l^4)$ term in the expansion of \eqn{anomglre}.


\end{document}